\documentclass[aps,prl,reprint,showpacs,superscriptaddress]{revtex4-1}

\usepackage{color}
\usepackage{graphicx}
\usepackage{amsmath}
\usepackage{amssymb}
\usepackage{epsfig}
\usepackage{dcolumn}
\usepackage{bm}

\begin{document}
\title{Electronic and photonic counting statistics\\ as probes of non-equilibrium quantum dynamics}
\date{June 7, 2016}

\author{Bj\"orn Kubala}
\affiliation{Institute for Complex Quantum Systems and IQST, Ulm University, Albert Einstein-Allee 11, 89069 Ulm, Germany}
\author{Joachim Ankerhold}
\affiliation{Institute for Complex Quantum Systems and IQST, Ulm University, Albert Einstein-Allee 11, 89069 Ulm, Germany}
\author{Andrew D. Armour}

\affiliation{School of Physics and Astronomy, University of Nottingham, Nottingham NG7 2RD, UK}

\begin{abstract}
When a mesoscopic conductor is coupled to a high-quality electromagnetic cavity the flow of charges and the flux of photons leaking out of the cavity can both depend strongly on the coupled quantum dynamics of the system. Using a voltage-biased Josephson junction as a model system, we demonstrate that there is a simple connection between the full counting statistics of the charges and the photons in the long time limit. We then reveal the intimate relation between the counting statistics and the nonlinear dynamics of the system, uncovering novel regimes of coherent charge {and} photon transport associated with bifurcations in the classical dynamics of the system.
\end{abstract}


\maketitle

\paragraph{Introduction.} The statistical properties of the radiation produced by mesoscopic electrical conductors can be very different to that produced by a classical conductor\,\cite{glauber,beenakker,lebedev}.  Such systems are also highly tunable since the radiation emitted depends not just on the conductor itself, but also on the properties of its electromagnetic environment and the details of the bias applied. Experiments over the last few years have been able to probe the photons emitted by a range of different conductors in unprecedented detail\,\cite{gabelli,hofheinz,portier,reulet} and their potential to act as sources of non-classical microwave radiation is being actively explored\,\cite{kubala2014,leppa:2015,mora,blais}.

\begin{figure}[t]
\centering
{\includegraphics[width=8.0cm]{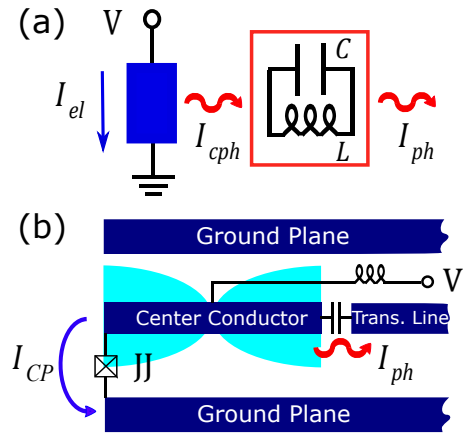}}
\caption{(Color online) (a) Schematic diagram of a voltage-biased mesoscopic conductor (blue box) coupled to an $LC$ oscillator (red box); an electrical current flowing through the conductor, $I_{el}$, generates a coherent flux of photons which enter the oscillator, $I_{cph}$, whilst the photons leaking out of the latter form another current, $I_{ph}$. (b) Josephson junction embedded in a superconducting cavity capacitively coupled to a transmission line. The dynamics of the system can be probed using either the Cooper-pair current in the circuit, $I_{CP}$, or the photons leaking out of the cavity into the transmission line, $I_{ph}$.}
\label{fig:schematic}
\end{figure}

Introducing an electromagnetic cavity into a circuit can transform the properties of both the electrical current and the radiation emitted\,\cite{holst,ingold}. Recent experiments using cavity-coupled Josephson junctions and semiconductor quantum dots, where the coupling was predominantly to a single cavity mode, showed that a strongly non-equilibrium photon population can be generated leading to novel regimes of strongly coupled charge and photon dynamics\,\cite{astafiev,chen:2014,petta:2015}. In such a system, illustrated schematically in Fig.\ \ref{fig:schematic}(a), the photon flux entering the mode is balanced by a leakage of radiation out into the wider electromagnetic environment and the  mode itself can be thought of as a conductor, albeit a photonic one rather than an electrical one. From this perspective it is natural to ask how the counting statistics of the photons flowing in and out of the mode are related to the charge current and what they tell us about the nonlinear quantum dynamics of the system. We address these questions in this Letter, focussing on a specific conductor, a voltage-biased Josephson junction coupled to a cavity, where the connection between the flow of an electrical current and the generation of photons is particularly simple and the coupled quantum dynamics especially rich.

When a Josephson junction (JJ) in series with a microwave cavity\,\cite{holst,ingold,hofheinz,pashkin,chen:2014} is biased at voltages below the gap energy, quasiparticle excitations are unable to dissipate energy. A dc current flows at resonances where the energy available to tunneling Cooper pairs  matches that required to generate one or more cavity photons. Since all of the energy from the voltage source is converted into photons, the dc current is strictly proportional to the rate at which photons are produced\,\cite{hofheinz,chen:2014}, in contrast to other conductors, such as semiconductor quantum dots, where typically only a small fraction of the charges generate photons\,\cite{petta:2015,xu}. Recent theoretical work has shown that the strong  nonlinearities in such systems can lead to non-classical features near resonances where each Cooper pair produces one or two photons\,\cite{paduraiu:12,leppa:13,armour2013,gramich2013,kubala2014,leppa:2015,leppa:2016}.

We begin by demonstrating a simple connection between the full counting statistics of the photonic and charge currents, not just their averages. We then examine the statistics of the charge and photon currents at the resonance where each Cooper pair generates one photon: We explore the emergence and eventual disappearance of a regime of strongly coherent transport\,\cite{grabertepl} as the Josephson energy is increased, signalled by strong suppression of the fluctuations in the charge and photon statistics, and establish its connection with a bifurcation that occurs in the classical dynamics of the cavity. The statistics are strikingly different to other systems where a mesoscopic conductor generates a non-equilibrium photon or phonon population\,\cite{novotny,flindt,nemsclerk,usmani,harvey,nemsfabio,lambert}, as well as to those at the two-photon resonance\,\cite{paduraiu:12}, where a bifurcation in the corresponding classical dynamics is instead typically marked by a strong {\emph{enhancement}} in the noise.

\paragraph{Josephson-cavity system.}\label{sec:model}
The model system we study consists of a JJ in series with an $LC$ oscillator to which a (sub-gap) voltage bias, $V$, is applied\,\cite{hofheinz,chen:2014}; a possible realization is shown in Fig.\ \ref{fig:schematic}(b). The oscillator is one of the modes of a high-Q superconducting microwave cavity which is assumed to be weakly coupled to a transmission line through which photons leak out of the system\,\cite{armour2013}.

The Hamiltonian of this JJ-oscillator system takes the time-dependent form $H=\hbar \omega_0 a^\dagger a -E_J \cos(\omega_J t +\varphi)$ \,\cite{armour2013,gramich2013}, where $a$ is the lowering operator for the oscillator which has frequency $\omega_0=1/\sqrt{LC}$, $E_J$ is the Josephson energy of the junction and $\omega_J=2eV/\hbar$ the Josephson frequency set by the bias voltage. The phase of the JJ is locked to the phase of the oscillator\,\cite{footnote} $\varphi=\Delta_0 (a^\dagger + a)$ with $\Delta_0=(2e^2 /\hbar)^{1/2}(L/C)^{1/4}$ the oscillator's zero point flux fluctuations in units of the flux quantum. While early experiments operated in the low impedance regime $\Delta_0\ll 1$\,\cite{hofheinz,chen:2014}, recent progress in circuit designs allows for $\Delta_0\sim O(1)$ \,\cite{larged1,larged2}.

We will focus on situations where the Josephson frequency is close to an integer, $p$, times the resonator frequency, $\omega_J\simeq p\omega_0$. In such cases, one can perform a rotating wave approximation (RWA) and obtain the effective Hamiltonian\,\cite{armour2013,gramich2013} (in a frame rotating at frequency $\omega_J/p$):
 \begin{eqnarray}
H^{(p)}_{\rm RWA}&=&\hbar\delta_{(p)} a^{\dagger}a \label{eq:hrwa}\\
&&-\frac{(-i)^p\tilde{E}_J}{2}:\left[\left(a^{\dagger}\right)^p+(-1)^p a^p\right]
\frac{{J}_p(2\Delta_0\sqrt{a^{\dagger}a})}{(a^{\dagger}a)^{p/2}}:, \nonumber
\end{eqnarray}
where the renormalized Josephson energy is defined as $\tilde{E}_J=E_J{\rm e}^{-\Delta_0^2/2}$, the detuning is given by $\delta_{(p)}=\omega_0-\omega_J/p$ and colons imply normal ordering. 

Including weak coupling between the oscillator and its surroundings\,\cite{armour2013,gramich2013} (i.e. the transmission line), assumed to be at zero temperature for simplicity, we can write down a master equation for the oscillator $\dot{\rho}=\mathcal{L}_{\chi=0}\,[\rho]$, with the Liouvillian
\begin{equation}
\mathcal{L}_\chi [\rho]=-\frac{i}{\hbar}[{H}_{\rm RWA}^{(p)},\rho]+\frac{\gamma}{2}\left(2 {\rm e}^{ -i\chi}\, a\rho a^{\dagger}-a^{\dagger}a\rho-\rho a^{\dagger}a\right), \label{eq:mas1}
\end{equation}
where $\gamma$ is the energy relaxation rate. The dissipative terms account for the irreversible loss of photons; the factor ${\rm e}^{ -i\chi}$ allows one to count the flow of these photons out of the oscillator\,\cite{gardiner,xu}.

\paragraph{Counting statistics of photons and charges.}
The \emph{photonic} current and its statistical properties are connected in a simple way with those of the dc Cooper-pair current flowing through the JJ, as we now demonstrate.
We  consider the generating function
$\mathcal{F}_{ph}(\chi,t) = \ln [\,\sum_{N} P(N, t) \,{\rm e}^{i \chi N}\,]$ of the number distribution $P(N, t)$ of photons leaking from the cavity during a time interval $t$. It is obtained by evolving the system starting from steady state, $\rho(0)$, with the Liouvillian \eqref{eq:mas1}  as
$\exp[{\mathcal{F}}_{ph}(\chi, t)]={\rm Tr}\{\exp[\mathcal{L}_\chi t]\, \rho(0)\}$.
The mean photon current and its moments follow from the cumulants\,\cite{belzig,kinderman,romito} $\kappa_{ph}^{(k)} = \langle\!\langle N^k \rangle\!\rangle = \left.\partial^k \mathcal{F}_\mathrm{ph}(\chi,t)/ \partial (i\chi)^k \right|_{\chi=0}\;$.
For long counting times $\mathcal{F}_{ph}(\chi, t\to \infty)\to t \lambda(\chi)$ with $\lambda$ the eigenvalue  of the counting Liouvillian $\mathcal{L}_\chi$ with the least negative real part.

In Liouville space (where the density operator turns into a vector), we can now introduce a unitary transformation,
$\exp[{\mathcal{F}}_{ph}(\chi, t)]={\rm Tr}\{ \mathcal{U}^{-1}   \exp[\mathcal{U} \mathcal{L}_\chi \mathcal{U}^{-1} t]\; \mathcal{U} \rho(0)\}$, where $ \left(\mathcal{U}  \rho(0)\right)_{nm} = \exp{[ -i (n+m) \chi/2]} \; \rho(0)_{nm}$.
Note, that $\mathcal{U}$ can not be cycled under the trace, which is to be taken in Hilbert space.
For the transformed Liouvillian, one finds 
 \begin{eqnarray}\label{eq:LCP}
 \mathcal{L}^{cph}_{p\chi} [\rho] &=& \mathcal{U} \mathcal{L}_\chi \mathcal{U}^{-1}\;  [\rho]\\
 &=& \frac{i}{\hbar} (H_{\chi}\rho-\rho H_{-\chi}) +\frac{\gamma}{2} (2 a \rho a^\dagger  -a^{\dagger}a\rho-\rho a^{\dagger}a)\, ,\nonumber
\end{eqnarray}
 where
 $H_\chi = \hbar\delta_{(p)} a^{\dagger}a  - (-i)^p (\tilde{E}_J/2) \; : [ {\rm e}^{-i p\chi/2} \left(a^{\dagger}\right)^p+(-1)^p \, {\rm e}^{i p\chi/2} a^p ]
J_p(2\Delta_0\sqrt{a^{\dagger}a})/(a^{\dagger}a)^{p/2}:$. This equation can also be interpreted as describing a counting process  in its own right. The particular way the counting field $p\chi$ appears in the coherent part indicates the quantity which is counted\,\cite{belzig,kinderman,romito}: namely packages of $p$ photons absorbed or emitted during a tunneling process.
Formally, this allows us to define a generating function ${\mathcal{F}}_{cph}(p \chi, t) := \ln \left( {\rm Tr}\{\exp[  \mathcal{L}^{cph}_{p\chi} t] \, \rho(0)\} \right)$ for a distribution $\tilde{P}(N,t)$ describing the coherent transfer of photons.
In  the long time limit, the behavior is dominated by $\lambda$ and hence
\begin{equation}\label{eq:FCS_eq}
{\mathcal{F}}_{cph}(p \chi, t)  =
{\mathcal{F}}_{ph}(\chi, t)
 \end{equation}
 up to an irrelevant constant, which immediately relates the cumulants $\kappa^{(k)}_{cph} = \kappa_{ph}^{(k)}/p^k$.

The  current operator for the coherent transfer of photons is then obtained by subtracting the absorption and emission terms for $p$ photons that appear in $H_{\chi}$,
 \begin{equation}
I_{cph}=\frac{i^{p-1}\tilde{E}_J}{2\hbar}:\frac{{J}_p(2\Delta_0\sqrt{a^{\dagger}a})}{(a^{\dagger}a)^{p/2}}
\left[a^p-(-1)^{p}(a^{\dagger})^p\right]:\ .
\end{equation}
However, this is equivalent to the operator describing the dc Cooper-pair current, the time-independent part of the standard Josephson current operator $I_J=\frac{2eE_J}{\hbar}\sin(\omega_Jt+\varphi)$  in the rotating frame and within the RWA:  $I_{CP}= 2eI_{cph}$. Hence in the long time limit, the statistics of Cooper pairs transferred through the JJ will match that of the photons leaving the cavity.

\begin{figure}[t]
\centering
{\includegraphics[width=8.5cm]{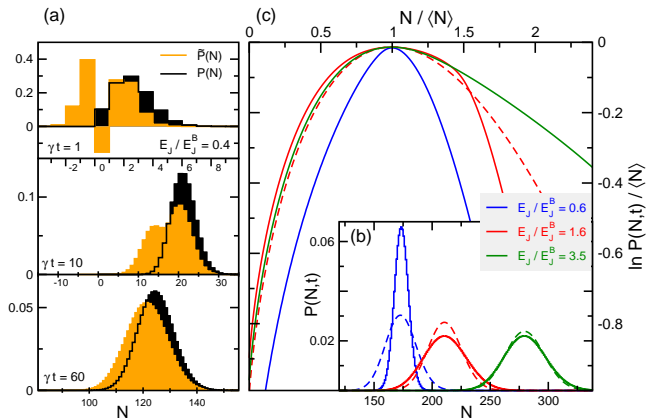}}
\caption{(Color online) (a) Distributions for coherent photon transfer into the cavity, $\tilde{P}(N,t)$, and incoherent emission from the cavity, $P(N,t)$, at different times. (b) Behavior of $P(N,t)$ below and above the bifurcation at $E_J=E_J^{B}$ together with the corresponding Poissonian distributions (dashed); $\gamma t=80$ (except for $E_J/E_J^{B}=1.6$ where $\gamma t=60$ for clarity).
(c) Large deviation functions (for $\gamma t=80$); the Poissonian case is shown as a dashed line. $\Delta_0=0.5$ throughout.}
\label{fig:fcs}
\end{figure}

We now look in detail at how the counting statistics reflect the rich nonlinear quantum dynamics of the system, focussing in the first instance on the case of the one photon resonance. Figure \ref{fig:fcs}(a) compares the evolutions of $P(N,t)$ and $\tilde{P}(N,t)$ over time (for details of the calculation see \cite{supplmat}). Although these distributions become identical in the limit of long times, the short time behavior is radically different. This is hardly surprising as $\tilde{P}(N,t)$ is a \emph{quasiprobability} distribution\,\cite{belzig,clerkfcs}, reflecting the coherence of the photon transfer from junction to cavity: not only can photons both enter ($N>0$) and leave the system ($N<0$), the precise number that have done so is ambiguous because of quantum coherences. This essential ambiguity is signalled by the negativity which is present in $\tilde{P}(N,t)$ for short times. It is dissipation which, over time, sets the direction for the flow of photons between the JJ and the cavity as well as destroying coherences so that $\tilde{P}(N,t)$ eventually becomes completely positive and converges with $P(N,t)$.

 The dynamics of the system becomes strongly nonlinear as $E_J$ is increased and this is reflected in the  shape of the distribution, $P(N,t)$ [and that of $\tilde{P}(N,t)$], that emerges in the limit of long times. The distribution is only ever Poissonian in the limit of very weak coupling (not shown); as $E_J$ is increased progressively $P(N,t)$ initially narrows dramatically before broadening again, becoming noticeably broader than the Poissonian case [see Fig.~\ref{fig:fcs}(b)]. The crossover from narrowing to broadening is linked with a bifurcation as we shall see below. The large deviation function for the photon statistics, Fig.\ \ref{fig:fcs}(c), shows that the behavior remains strongly non-Poissonian over the whole range of $E_J$ studied.

\paragraph{Charge and photon Fano factors.}
To examine the counting statistics around the bifurcation in more detail we now analyze the behavior of the second cumulants. These quantities can clearly be derived directly from the full counting distributions (\ref{eq:FCS_eq}), but they can also be obtained from the behavior of second order correlations functions. This latter approach is advantageous in that it provides a simple way of applying analytic techniques such as the semiclassical approximation which in turn provides insight into the relation between noise properties and nonlinear quantum dynamics.

Accordingly, we consider for both Cooper pair ($\mu=CP$) and photonic ($\mu=ph$) current, respectively, $S_{\mu}=2\int_0^{\infty} d t \left[\langle I_{\mu}(t) I_{\mu}(0)\rangle-\langle I_{\mu}\rangle^2\right]$.
In particular, the Fano factors $F_{\mu}=S_{\mu}/(q_\mu \langle I_{\mu}\rangle)$ with $q_{CP}=2 e$, $q_{ph}=1$ provide a convenient way of quantifying deviations from the Poissonian value ($F_{\mu}=1$).
Using the general relation (\ref{eq:FCS_eq}), one then finds  that $F_{ph}=p\, F_{CP}$
which can be considered a generalization of the Ramo-Shockley relation in quantum transport\,\cite{rs,flindt}.

In the semiclassical approximation the basic idea is to expand about the classical fixed points of the system linearizing the equations of motion for the corresponding quantum fluctuations\,\cite{armour2013} $a\to \alpha+\delta a$. This applies to the regime where $\Delta_0\ll 1$ and photon numbers become sufficiently large. Formally, one moves to a displaced frame using $D(\alpha)={\rm e}^{\alpha a^{\dagger}-\alpha^*a}$ with $\alpha$ a complex amplitude. Quantum noise properties  are now captured by a master equation as in (\ref{eq:mas1}) with the quadratic Hamiltonian
\begin{equation}
H_2^{(p)}=\hbar[\delta_{(p)}+\nu_{(p)}]\delta a^{\dagger}\delta a+i\frac{\hbar}{2}\left(g_{(p)}\delta a^{\dagger}\delta a^{\dagger}-g^*_{(p)}\delta a\delta a\right) \label{eq:h2}
\end{equation}
where the coefficients $\nu_{(p)}(A,\phi), g_{(p)}(A,\phi)$ depend on amplitude and phase of the classical fixed points $\alpha= A {\rm e}^{i\phi}$ (see \cite{supplmat}). The stability of these fixed points as well as the dynamics of the correlation functions is then determined by the eigenvalues of the matrix of the equations of motion for the expectation values of $\delta a$, $\delta a^{\dagger}$, i.e.,
\begin{equation}
\Gamma_{(p)}^{\pm}=-\frac{\gamma}{2}\pm\sqrt{|g_{(p)}|^2-(\delta_{(p)}+\nu_{(p)})^2}. \label{eq:eigs}
\end{equation}
On-resonance, as $E_J$ is increased from zero both eigenvalues are initially real and finite until one arrives at bifurcations, where one of them vanishes. For sufficiently large $E_J$,   eventually $\Gamma_{(p)}^{\pm}$ become complex.

\begin{figure}[t]
\centering
\includegraphics[width=7.0cm]{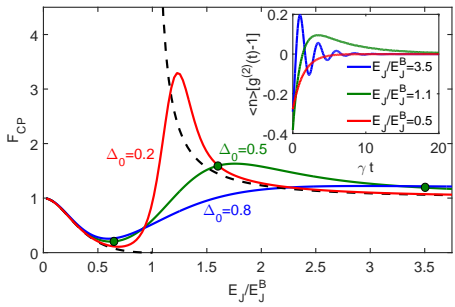}
\caption{(Color online) Cooper-pair Fano factor, $F_{CP}$, at the one-photon resonance as a function of $E_J/E_J^B$ with $\delta_{(1)}=0$. Full lines are numerical results using $I_{CP}$ and the full quantum dynamics, dashed lines are from the semiclassical calculation below and above the bifurcation. Inset: Corresponding behavior of $\langle a^{\dagger}a\rangle[g^{(2)}(t)-1]$ as a function of $t$  for $\Delta_0=0.2$. Green dots on the curve for $\Delta_0=0.5$ indicate the $E_J/E_J^B$ values for the distributions in Figs.\ \ref{fig:fcs} (b) and (c).}
\label{fig:p1fanoc}
\end{figure}

\paragraph{One-photon resonance.}
The classical system undergoes a bifurcation at $E_J=E_J^{B}=\hbar\gamma z_1{\rm e}^{\Delta_0^2/2}/[4J_0(z_1)\Delta_0^2]$ with $z_1= 1.841$ the first maximum of the Bessel function $J_1(z)$\,\cite{armour2013,gramich2013}. Below the bifurcation there is a single stable fixed point whose amplitude, $A$, grows with $E_J$.

Using the semiclassical approximation, below the bifurcation ($E_J<E_J^B$) and on-resonance one finds
\begin{equation}
F_{CP}=\left[\frac{dJ_1(z)/dz}{Y_1(z)}\right]^2\frac{Y_0(z_1)^2}{\left[Y_0(z_1)+\frac{E_J}{E_J^B}\, Y_2(z)\right]^2}, \label{eq:cnspec2}
\end{equation}
with $Y_k(z)=J_k(z)/z$ and $z=2\Delta_0A$. In the limit $E_J\rightarrow 0$,  the amplitude $A\rightarrow 0$ so also $F_{CP}\rightarrow 1$ and the Poissonian result is recovered\,\cite{grabertepl,gramich2013} (as discussed above) implying incoherent Cooper pair tunneling. However, as $E_J$ is increased the value of $F_{CP}$ drops, signifying a transition to a regime of increasingly coherent charge transport\,\cite{grabertepl}. Remarkably, the semiclassical calculation predicts that $F_{CP}\rightarrow 0$ as $E_J/E_J^B\rightarrow 1$.  Above the bifurcation is different: The semiclassical fluctuations in the current do not vanish as $E_J/E_J^B\rightarrow 1$ {\emph{from above}}, here the behavior is dominated by the emergence of a soft mode since one of the eigenvalues, $\Gamma^{\pm}_{(1)}$, tends to zero which results in a divergence of the current noise.

Quantum fluctuations smear out the discontinuity at the bifurcation as shown  in Fig.\ \ref{fig:p1fanoc} for a range of values of $\Delta_0$. A smooth drop in $F_{CP}$ as $E_J$ is increased is seen in the numerical results, but it is weaker than in the semiclassical limit  and the semiclassical discontinuity at the bifurcation is  echoed by a peak in $F_{CP}$. These features match the crossover from narrowing to broadening of the full distributions in Fig.\ \ref{fig:fcs}. However, for larger values of $\Delta_0$ the peak at the bifurcation is smoothed out and eventually disappears.

Well beyond the bifurcation the Fano factor of the Cooper-pair current noise tends again towards unity. In the semiclassical formalism this behavior follows from the fact that the eigenvalues of the fluctuations become complex implying that  two-time correlation functions, such as $g^{(2)}(t)=\langle a^{\dagger} a^{\dagger}(t)a(t)a\rangle/\langle a^{\dagger} a\rangle^2$, oscillate strongly\,\cite{carmichael:book} (see Fig.\ \ref{fig:p1fanoc}, inset).
The effect on the Fano factor can be seen clearly using the relation\,\cite{emary,xu}:
\begin{equation}
F_{ph}=1+2\gamma\langle a^{\dagger}a\rangle\int_0^{\infty} dt [g^{(2)}(t)-1] \label{eq:fphg}
\end{equation}
which implies $F_{ph}\rightarrow 1$ whenever $g^{(2)}(t)$ oscillates sufficiently strongly. Nevertheless, as we have seen in Fig.\ \ref{fig:fcs} the full current statistics remains non-Poissonian.

 \paragraph{Two-photon resonance.}
Rich quantum dynamics also appears for the two photon ($p=2$) resonance.
In this case there are {\em two} bifurcations as a function of the pump parameter $E_J$\,\cite{armour2013}: One corresponds to a parametric resonance with a threshold at
 ${E}_J^C=\hbar\gamma {\rm e}^{\Delta_0^2/2}/\Delta_0^2$, another at $E_J^{B2}=\hbar\gamma{\rm e}^{\Delta_0^2/2}/[4Y_1(z_2)\Delta_0^2]>E_J^C$ with $z_2=3.054$ to a bifurcation similar to that which occurs for $p=1$. Below $E_J^C$ a quadratic approximation of the Hamiltonian leads to\,\cite{supplmat}
\begin{figure}[t]
\centering
{\includegraphics[width=7.0cm]{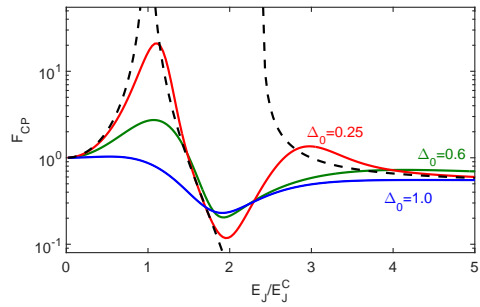}
}
\caption{(Color online) Cooper-pair Fano factor, $F_{CP}$, for the two-photon resonance as a function of $E_J/E_J^C$. Numerical results using $I_{CP}$ and the full quantum dynamics (full lines) are compared with the semiclassical and quadratic approximations (dashed lines).  Here $E_J^{B2}=2.4E_C$ and  $\delta_{(2)}=0$.}
\label{fig:fano2}
\end{figure}
$F_{CP}=(2+x^2+x^4)/[2\left(1-x^2\right)^2]$ with $x=E_J/E_J^C$. This starts, as expected, at unity in the small-$E_J$ limit, but in this case $F_{CP}$ {\em grows} with $E_J$ and diverges as the threshold is approached. Since in this case $F_{CP}=F_{ph}/2$, this matches the result for $F_{ph}$ in Ref.\ \onlinecite{paduraiu:12} which analyzed the sub-threshold photon counting statistics.
However, above threshold ($E_J\rightarrow  E_J^{B2}$)  a coherent regime emerges with  $F_{CP}<1$ before the second bifurcation is reached. Again strong quantum fluctuations erase the signatures of the classical bifurcations: for $\Delta_0=1.0$ the peak at the parametric threshold is converted into a smooth dip, as can be seen in Fig.\ \ref{fig:fano2}.

\paragraph{Conclusions.} Considering  a mesoscopic conductor coupled to a cavity, we have shown that a simple relation emerges in the long time limit between the full counting statistics of the charges and photons, not just their average currents. The statistics of the charges and photons depend strongly on the underlying coupled quantum dynamics, for the cavity-JJ system this leads to the emergence of novel regimes of coherent transport of Cooper pairs \emph{and} photons. This contrasts strongly with the behavior typically associated with bifurcations in systems where conductors excite non-equilibrium populations of photons (or phonons)\,\cite{novotny,flindt,nemsclerk,usmani,harvey,nemsfabio,lambert}. These results can be tested experimentally, given very recent progress in inferring discrete microwave statistics using continuous measurements\,\cite{reuletn}. Our work also raises a number of questions about how the counting statistics of different entities are linked, not just for cavity-conductor set-ups, but also optomechanical devices\,\cite{optomech} and other hybrid systems. These include the relation between the short time dynamics of the counting statistics and finite frequency noise, the connection between the coherent photon flow and work statistics\,\cite{work,pekola}, and the detection process for charge counting in general.

\paragraph{Acknowledgements.}
The authors thank F. Portier and S. Dambach for valuable discussions. The work was facilitated by support from the International Collaboration Fund of the University of Nottingham, BK and JA were also supported by Deutsche Forschungsgemeinschaft through AN336/6-1 and SFB/TRR21.

\bibliographystyle{apsrev4-1}

\begin{thebibliography}{99}
\bibitem{glauber} R. J. Glauber, Phys. Rev. {\bf 131}, 2766 (1963).
\bibitem{beenakker} C. W. J. Beenakker and H. Schomerus, Phys. Rev. Lett. {\bf 86}, 700 (2001).
\bibitem{lebedev} A. V. Lebedev, G. B. Lesovik, and G. Blatter, Phys. Rev. B {\bf 81}, 155421 (2010).
\bibitem{gabelli} J. Gabelli, L.-H. Reydellet, G. F\`{e}ve, J.-M. Berroir, B. Pla\c{c}ais, P. Roche, and D. C. Glattli, Phys. Rev. Lett. {\bf 93}, 056801 (2004).
\bibitem{hofheinz} M. Hofheinz, F. Portier, Q. Baudouin, P. Joyez, D. Vion, P. Bertet, P. Roche, and D. Esteve, Phys. Rev. Lett. {\bf 106}, 217005 (2011).
\bibitem{portier}  E. Zakka-Bajjani, J. Dufouleur, N. Coulombel, P Roche, D. C. Glattli and F. Portier, Phys. Rev. Lett. {\bf 104}, 206802 (2010).
\bibitem{reulet} J.-C. Forgues, C. Lupien and B. Reulet, Phys. Rev. Lett., {\bf 113}, 043602 (2014); Phys. Rev. Lett., {\bf 114}, 130403 (2015).
\bibitem{kubala2014}  B. Kubala, V. Gramich and  J. Ankerhold, Phys. Scr. {\bf T165},  014029 (2015).
\bibitem{leppa:2015} J. Lepp\"{a}kangas, M. Fogelstr\"{o}m, A. Grimm, M. Hofheinz, M. Marthaler and G. Johansson,
Phys. Rev. Lett. {\bf 115}, 027004 (2015).
\bibitem{mora} U. C. Mendes and C. Mora New Journal of Physics {\bf 17}, 113014 (2015).
\bibitem{blais} A. L. Grimsmo, F. Qassemi, B. Reulet, and A. Blais, Phys. Rev. Lett. {\bf 116}, 043602 (2016).
\bibitem{holst} T. Holst, D. Esteve, C. Urbina, and M. H. Devoret, Phys. Rev. Lett. {\bf 73}, 3455 (1994).
\bibitem{ingold} G.-L. Ingold and Y. V. Nazarov, in {\it Single Charge Tunneling},
edited by H. Grabert and M. H. Devoret (Plenum, New York, 1992).
\bibitem{astafiev} O. Astafiev, K. Inomata, A. O. Niskanen, T. Yamamoto, Yu. A. Pashkin, Y. Nakamura and J. S. Tsai, Nature (London) {\bf 449} 588 (2007).
\bibitem{chen:2014} F. Chen, J. Li, A. D. Armour, E. Brahimi, J. Stettenheim, A. J. Sirois, R. W. Simmonds, M. P. Blencowe and A. J. Rimberg, Phys. Rev. B {\bf 90}, 020506 (2014).
\bibitem{petta:2015} Y.-Y. Liu, J. Stehlik, C. Eichler, M. J. Gullans, J. M. Taylor and J. R. Petta, Science {\bf 347}, 285 (2015).

\bibitem{pashkin}Y. A. Pashkin, H. Im, J. Lepp\"{a}kangas, T. F. Li, O. Astafiev,
A. A. Abdumalikov, E. Thuneberg, and J. S. Tsai, Phys. Rev. B {\bf 83}, 020502 (2011).
\bibitem{xu} C. Xu and M. G. Vavilov, Phys. Rev. B {\bf 88}, 195307 (2013).


\bibitem{paduraiu:12} C. Padurariu, F. Hassler and Yu. V. Nazarov, Phys. Rev. B {\bf 86}, 054514 (2012).

\bibitem{leppa:13} J. Lepp\"{a}kangas, G. Johansson, M. Marthaler and M. Fogelstr\"{o}m,  Phys. Rev. Lett. {\bf 110}, 267004 (2013).
\bibitem{armour2013} A. D. Armour, M. P. Blencowe, E. Brahimi and A. J. Rimberg, Phys. Rev. Lett. {\bf 111}  247001 (2013).
\bibitem{gramich2013} V. Gramich, B. Kubala, S. Rohrer and J. Ankerhold, Phys. Rev. Lett. {\bf 111}  247002 (2013).

\bibitem{leppa:2016} J. Lepp\"{a}kangas, M. Fogelstr\"{o}m, M. Marthaler and G. Johansson,   Phys. Rev. B {\bf 93}, 014506 (2016).

\bibitem{grabertepl} H. Grabert and G.-L. Ingold, Europhys. Lett. {\bf 58} 429 (2002).

\bibitem{novotny} T. Novotn\'{y}, A. Donarini and A.-P. Jauho, Phys. Rev. Lett. {\bf 90}, 256801 (2003).
\bibitem{flindt} C. Flindt, T. Novotn\'{y} and A.-P. Jauho, Phys. Rev. B {\bf 70}, 205334 (2004).

\bibitem{nemsclerk} S. D. Bennett and A. A. Clerk, Phys. Rev. B {\bf 74}, 201301 (2006).
\bibitem{usmani} O. Usmani, Ya. M. Blanter and Yu. V. Nazarov, Phys. Rev. B {\bf 75}, 195312 (2007).
\bibitem{harvey} T. J. Harvey, D. A. Rodrigues and A. D. Armour, Phys. Rev. B {\bf 78}, 024513 (2008).
\bibitem{nemsfabio} J. Br\"{u}ggemann, G. Weick, F. Pistolesi, and F. von Oppen,
Phys. Rev. B {\bf 85} 125441 (2012).
\bibitem{lambert} N. Lambert, F. Nori and C. Flindt
Phys. Rev. Lett. {\bf 115}, 216803 (2015).
\bibitem{footnote} This treatment can also be generalised to include the effects of low frequency voltage fluctuations\,\cite{gramich2013}.

\bibitem{larged1} M. A. Castellanos-Beltran and K. W. Lehnert, Appl. Phys. Lett. {\bf 91}, 083509 (2007).
\bibitem{larged2}  C. Altimiras, O. Parlavecchio, P. Joyez, D. Vion, P. Roche, D. Esteve and F. Portier, Phys. Rev. Lett. {\bf 112}, 236803 (2014).
\bibitem{gardiner} C. W. Gardiner and P. Zoller, {\it Quantum Noise}, (Springer, Berlin, Germany 2004).

\bibitem{supplmat} See additional information provided in the Supplemental Material.
\bibitem{belzig} W. Belzig and Yu. V. Nazarov, Phys. Rev. Lett. {\bf 87}, 197006 (2001).
\bibitem{kinderman} Yu. V. Nazarov and M. Kinderman, Eur. Phys. J. B {\bf 35} 413 (2003).
\bibitem{romito} A. Romito and Yu. V. Nazarov, Phys. Rev. B {\bf 70}, 212509 (2004).
\bibitem{clerkfcs} P. P. Hofer and A. A. Clerk, Phys. Rev. Lett. {\bf 116}, 013603 (2016).

\bibitem{rs} J. H. Davies, P. Hyldgaard, S. Hershfield and J. W. Wilkins, Phys. Rev. B {\bf 46}, 9620 (1992).
\bibitem{emary} C. Emary, C P\"{o}ltl, A. Carmele, J. Kabuss, A. Knorr and T. Brandes, Phys. Rev. B {\bf 85}, 165417 (2012).
\bibitem{carmichael:book} H. J. Carmichael, {\it Statistical Methods in Quantum Optics, Vol. 1} (Springer, Heidelberg, 1999).
\bibitem{reuletn} S. Virally, J. O. Simoneau, C. Lupien and B. Reulet, Phys. Rev. A {\bf 93}, 043813 (2016).
\bibitem{optomech} M. Aspelmeyer, T. J. Kippenberg and F. Marquardt,
Rev. Mod. Phys. {\bf 86} 1391 (2014).
\bibitem{work} M. Esposito, U. Harbola and S. Mukamel, Rev. Mod. Phys. {\bf 81}, 1665 (2009).
\bibitem{pekola} J. Pekola, Nature Phys. {\bf 11}, 118 (2015).
\end{thebibliography}

\begin{thebibliography}{99}
\bibitem{scarmichael:book} H. J. Carmichael, {\it Statistical Methods in Quantum Optics, Vol. 1} (Springer, Heidelberg, 1999).
\bibitem{sxu} C. Xu and M. G. Vavilov, Phys. Rev. B {\bf 88}, 195307 (2013).
\bibitem{sarmour2013} A. D. Armour, M. P. Blencowe, E. Brahimi and A. J. Rimberg, Phys. Rev. Lett. {\bf 111}  247001 (2013).

\end{thebibliography}

\pagebreak
~

\pagebreak

\pagebreak

\onecolumngrid
\vspace{\columnsep}
\begin{center}
\textbf{\large Electronic and photonic counting statistics\\ as probes of non-equilibrium quantum dynamics: Supplemental material}
\end{center}
\vspace{\columnsep}
\twocolumngrid

\setcounter{equation}{0}
\setcounter{figure}{0}
\setcounter{table}{0}
\setcounter{page}{1}
\makeatletter
\renewcommand{\theequation}{S\arabic{equation}}
\renewcommand{\thefigure}{S\arabic{figure}}

\newcommand\id{\ensuremath{\mathbbm{1}}}

\section*{Numerical calculation of photon counting distributions $P(N,t)$ and $\tilde{P}(N,t)$}

In Fig. 2 of the main text we show numerical results for the distributions $P(N,t)$ and $\tilde{P}(N,t)$ for the one-photon ($p=1$) resonance. These are defined from the corresponding generating functions
\begin{eqnarray}
\mathcal{F}_\mathrm{ph}(\chi,t) &=& \ln [\,\sum_{N=0}^\infty P(N, t) \,{\rm e}^{i \chi N}\,] \\
\mathcal{F}_{cph}(p \chi, t) &=&  \ln [\,\sum_{N=-\infty}^\infty \tilde{P}(N, t) \,{\rm e}^{i \chi N}\,] \label{eq:S2}
\end{eqnarray}
via inverting the Fourier transformation.
The generating functions for counting either coherent photon flow or the photonic leakage from the cavity follow in turn from time evolution with the corresponding Liouvillians (Eqs. (2) and (3) of the main text), which contain the counting field in the coherent or the dissipative part of the Liouvillian. In that manner the order $p$ of the $p$-photon resonance enters into the generating function for coherent photon flow, $\mathcal{F}_{cph}(p \chi, t)$, and thus also into the $\tilde{P}(N, t)$ defined by \eqref{eq:S2}. $\tilde{P}$ can also be rewritten as counting packages of $p$ photons by using the variable $M=N/p$ in which case the exponential factor in \eqref{eq:S2} becomes ${\rm e}^{ip\chi M}$.
Note, that $\tilde{P}(N,t)$ has contributions from positive and negative values of $N$ as the generating function $\mathcal{F}_{cph}(p \chi, t)$ contains powers of ${\rm e}^{\pm i p\chi/2}$ while $P(N,t)$ only has contributions for $N\ge 0$ as we only consider leakage from the cavity (zero-temperature limit).

Numerically, we obtain the quasi-probabilty distribution $\tilde{P}(N,t)$ exactly following these definitions:\\
A numerical representation of ${\mathcal{F}}_{cph}(p \chi, t)$ for a fixed time $t$ is calculated from time-evolution with $\mathcal{L}^{cph}_{p\chi}$, where a sufficient number of $p\chi \in [0, 4 \pi]$ values  are sampled from the full period of ${\mathcal{F}}_{cph}(p \chi, t)$ (which is a function of $p\chi/2$). Then the Fourier-integral is numerically evaluated for various values of $N$.

The probabilty distribution $P(N,t)$ can be calculated in an analogous manner (with the time-evolution now governed by $\mathcal{L}_\chi$ instead of  $\mathcal{L}^{cph}_{p\chi}$). Alternatively, we employed for larger times the $N$-resolved density matrix approach \cite{scarmichael:book,sxu}, where a set of density matrices $\rho^{(N)}(t)$ representing the density matrix of the system after $N$ photons have left the cavity is evolved according to
\begin{equation}\label{eq:n-resolved}
\dot{\rho}^{(N)}(t) =   \mathcal{L}^\mathrm{det} [\rho^{(N)}(t)] +   \mathcal{J} [\rho^{(N-1)}(t)] \,,
 \end{equation}
where  $\mathcal{L}_{\chi=0} =   \mathcal{L}^\mathrm{det}  +   \mathcal{J} $ with $ \mathcal{J}[\rho] = \gamma a \rho a^\dagger$.

The probabilty distribution then simply follows as $P(N,t) = {\rm Tr}\{\rho^{(N)}(t)\}$.

\section*{Semiclassical and quadratic approximations}
The quantities $\nu_{(p)}$ and $g_{(p)}$ that appear in the Hamiltonian that arises within the semiclassical approximation [Eq.\ (6)] are defined as follows:
\begin{eqnarray}
\nu_{(p)}(A,\phi)&=&\frac{\tilde{E}_J\Delta_0^2}{\hbar} J_p(2\Delta_0A)\cos[p(\phi-\pi/2)]\label{eq:nup}\\
g_{(p)}(A,\phi)&=&-i\frac{\tilde{E}_J\Delta_0^2}{2\hbar}\left\{J_{p-2}(2\Delta_0A){\rm e}^{i(p-2)(\phi-\pi/2)}\right.\nonumber \\
&&\left.+J_{p+2}(2\Delta_0A){\rm e}^{-i(p+2)(\phi-\pi/2)}\right\},
\end{eqnarray}
with $A$ and $\phi$ the amplitude and phase of the classical fixed point\,\cite{sarmour2013}.

When the fixed point amplitude is non-zero, the current noise is calculated in the semiclassical approximation using just the linear current fluctuations
\begin{equation}
\delta I=K_{(p)}\delta a+K_{(p)}^*\delta a^{\dagger}, \label{eq:di}
\end{equation}
where $\delta I=I_{CP}-\langle I_{CP}\rangle$ with
\begin{eqnarray}
K_{(p)}&=&\frac{i^{p-1}e\tilde{E}_J\Delta_0}{\hbar}\left[J_{p-1}(2\Delta_0A_0){\rm e}^{-i\phi_0(p-1)}\right. \nonumber\\
&&\left.+(-1)^pJ_{p+1}(2\Delta_0A_0){\rm e}^{i\phi_0(p+1)}\right].
\end{eqnarray}
The correlation functions for the current fluctuations can be calculated using the regression formula\,\cite{scarmichael:book}, leading to a
simple expression for the current noise
\begin{equation}
S_{CP}=\frac{\gamma\left|K_{(p)}\left[i(\nu_{(p)}+\delta_{(p)})-\gamma/2\right]-K^*_{(p)}g_{(p)}\right|^2}{\Gamma^+_{(p)}\Gamma^-_{(p)}}. \label{eq:s0main}
\end{equation}
The denominator in this expression is written in terms of the eigenvalues that describe the evolution of the linear fluctuations, $\Gamma_{(p)}^{\pm}$ [see Eq.\ (7)]. This connection is important for understanding the behavior in the vicinity of the bifurcations which the system undergoes: one of the eigenvalues vanishes at these points, implying a divergence in the noise provided the denominator remains non-zero.

In cases where $A=0$, such as the two-photon resonance below threshold ($E_J<E_J^{C}$) this linear approach is not sufficient.
 In this case there is no displacement transformation and one simply uses quadratic approximations for the Hamiltonian \emph{and} the current operator (i.e.\ $\delta a=a$).
In the quadratic approximation the average current and occupation number $n=a^{\dagger}a$ operator obey the coupled equations
\begin{eqnarray}
\langle \dot{I}_{CP}^{(2)}\rangle&=&-\gamma\langle {I}_{CP}^{(2)}\rangle+\frac{e\tilde{E}_J^2\Delta_0^4}{2\hbar^2}\left[\langle n\rangle+1/2\right]\\
\langle \dot{n}\rangle&=&-\gamma\langle n\rangle+\langle{I}_{CP}^{(2)}\rangle/e,
\end{eqnarray}
where the quadratic Cooper pair current operator is
\begin{equation}
{I}_{CP}^{(2)}=\frac{ie\tilde{E}_J\Delta_0^2}{2\hbar}(aa-a^{\dagger}a^{\dagger}).
\end{equation}
The two-time correlation function for the current is then again obtained using the regression formula.

~

\end{document}